\documentclass[a4paper,11pt]{article}
\pdfoutput=1 

\usepackage{jheppub} 

\usepackage[T1]{fontenc} 

\title{Measurement of the forward $\eta$ meson production rate in p-p collisions at $\sqrt{s}$=13 TeV with the LHCf-Arm2 detector}

\author[a,b]{O. Adriani,}
\affiliation[a]{INFN Section of Firenze, Italy
}
\affiliation[b]{University of Firenze, Italy}
\author[a]{E. Berti,}
\author[a,b]{P. Betti,}
\author[a]{L. Bonechi,}
\author[a,b]{M. Bongi,}
\author[a,b]{R. D'Alessandro,}
\author[a]{S. Detti,}
\author[c]{M. Hagenauer,}
\affiliation[c]{Ecole-Polytechnique, Palaiseau, France
}
\author[d,e]{Y. Itow,}
\affiliation[d]{Institute for Space-Earth Environmental Research, Nagoya University, Nagoya, Japan}
\affiliation[e]{Kobayashi-Maskawa Institute for the Origin of Particles and the Universe, Nagoya University, Nagoya, Japan}
\author[f]{K. Kasahara,}
\affiliation[f]{Faculty of System Engineering, Shibaura Institute of Technology, Japan
}
\author[d]{Y. Kitagami,}
\author[d]{M. Kondo,}
\author[d]{Y. Matsubara,}
\author[d]{H. Menjo,}
\author[d]{Y. Muraki,}
\author[d]{K. Ohashi,}
\author[a]{P. Papini,}
\author[g,h,i,1]{G. Piparo,\note{Corresponding author.}}
\affiliation[g]{INFN Section of Catania, Italy }
\affiliation[h]{University of Catania, Italy}
\affiliation[i]{CSFNSM, Catania, Italy}
\author[a,j]{S. Ricciarini,}
\affiliation[j]{IFAC-CNR, Italy}
\author[k]{T. Sako,}
\affiliation[k]{ICRR, University of Tokyo, Kashiwa, Japan
}
\author[l]{N. Sakurai,}
\affiliation[l]{Tokushima University, Tokushima, Japan
}
\author[a]{M. Scaringella,}
\author[m]{Y. Shimizu,}
\affiliation[m]{Kanagawa University, Kanagawa, Japan
}
\author[m]{T. Tamura,}
\author[a,b]{A. Tiberio,}
\author[n]{S. Torii,}
\affiliation[n]{RISE, Waseda University, Shinjuku, Tokyo, Japan
}
\author[g,h,i]{A. Tricomi,}
\author[o]{W.C. Turner}
\affiliation[o]{LBNL, Berkeley, California, USA 
}
\author[f]{and K. Yoshida}

\emailAdd{giuseppe.piparo@ct.infn.it, giuseppe.piparo@cern.ch}

\abstract{The forward $\eta$ mesons production has been observed by the Large Hadron Collider forward (LHCf) experiment in proton-proton collision at $\sqrt{s}$=13 TeV. This paper presents the measurement of the inclusive production rate of $\eta$ in $p_T<$ 1.1 GeV/c, expressed as a function of the Feynman-x variable. These results are compared with the predictions of several hadronic interaction models commonly used for the modelling of the air showers produced by ultra-high energy cosmic rays. This is both the first measurement of $\eta$ mesons from LHCf and the first time a particle containing strange quarks has been observed in the forward region for high-energy collisions. These results will provide a powerful constraint on hadronic interaction models for the purpose of improving the understanding of the processes underlying the air showers produced in the Earth's atmosphere by ultra-energetic cosmic rays.}

\begin{document} 
\maketitle
\flushbottom

\section{\label{sec:level1}Introduction}
Due to the steeply decreasing flux, ultra-high energy cosmic rays (UHECRs) can be measured only indirectly by observing the air showers induced in the Earth's atmosphere. Many ground-based experiments performed measurements of flux, composition and anisotropy of UHECRs \cite{abraham2008observation,abreu2010update,abraham2010measurement,abbasi2009measurement,abbasi2010analysis,abbasi2010indications,sagawa2011recent}. Recently the Pierre Auger Observatory (PAO \cite{pierre2015pierre}) and Telescope Array (TA \cite{fukushima2003telescope}) experiments made the most precise measurements using hybrid detection techniques. The information about the primary cosmic rays is obtained by inferring the air shower characteristics from an array of surface detectors to observe charged particles at ground and a stereoscopic system of fluorescence detectors to observe light emission due to the excitation and de-excitation of molecules in the atmosphere. Despite the progress made through combining these techniques, the interpretation of the results obtained on mass composition still needs to be clarified \cite{aab2014depth,aab2014muons,abbasi2015study}. This is because critical parts of the data analyses depend on the simulation of the air showers. The hadronic interaction models used to predict the consecutive elementary interactions during the shower development are a fundamental input in the simulations. Models must provide predictions on both "hard" and "soft" processes. While the Standard Model properly describes hard processes, soft processes need phenomenological treatment because of the low transferred four-momentum, which forbids the development of a perturbation theory because of the high values of the QCD coupling constant. Soft processes can be described using Gribov-Regge theories \cite{gribov1968reggeon, regge1959introduction}, which involve the exchange of virtual quasi-particles called "Pomerons" in the interactions. Since there are different ways to implement Gribov-Regge theories, various models based on different implementations provide conflicting results on particle production. In order to reduce the differences between models, it is necessary to have high-energy calibration data. The Large Hadron Collider (LHC \cite{evans2008lhc}) is the most suitable place to perform these measurements since a proton-proton collision at $\sqrt{s}=13$ TeV is equivalent to the interaction of about $10^{17}$ eV cosmic ray with a proton at rest. Indeed, many experiments have provided calibration data for hadronic interaction models in recent years  (see \cite{albrecht2022muon} for a review of the most up-to-date results). Among these, the Large Hadron Collider forward (LHCf \cite{adriani2008lhcf}) experiment was designed to measure the production of neutral particles at very high pseudorapidities, which is one of the fundamental parameters of the models, since soft processes are mainly associated with the forward particle production. This paper reports the first LHCf measurement of $\eta$ meson inclusive production rate, which was accomplished using the LHCf-Arm2 detector. The importance of this observation relies on the fact that $\eta$ meson is a probe for the contribution of strange quarks to the hadronization mechanism. Differences in this parameter induce a large discrepancy in the expected $\eta$ production cross section among the models \cite{adriani2019lhcf}. In addition, the photons produced in the $\eta$ decays are the second dominant photon sources in high energy hadronic interactions after $\pi^{0}$ production and are relevant for the development of air showers. This paper is organized as follows. Section \ref{sec:lhcf_exp} describes the LHCf experimental apparatus. In Section \ref{sec:cond}, we summarize the experimental data set and the Monte Carlo (MC) simulation methodology. Section \ref{sec:ana_frame} explains the general analysis strategy, including the reconstruction procedure, the selection criteria, the calculation of corrections for the final spectrum and the effects contributing to the systematic uncertainty. In Section \ref{sec:res}, the analysis results are presented and compared with the predictions of several hadronic interaction models. Finally, the paper conclusions can be found in Section \ref{sec:concl}.
\section{\label{sec:lhcf_exp}The LHCf experiment}
\begin{figure}[]
\centering
\includegraphics[width=0.8\linewidth]{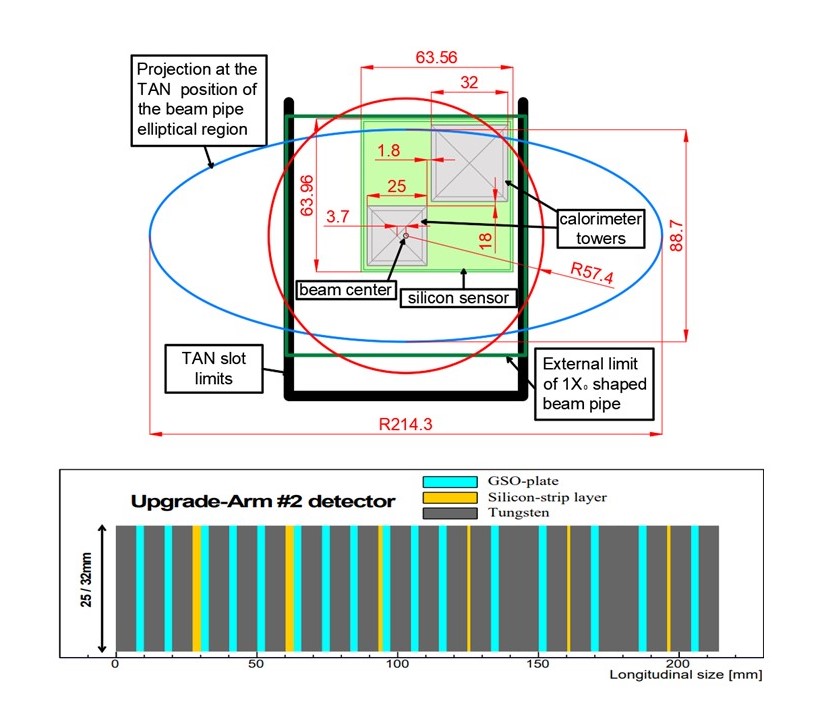}
\caption{\label{fig:lhcf_det}Schematic view of the LHCf-Arm2 detector. The transversal view of the detector inside the TAN slot is illustrated in the top panel, while the bottom panel shows the longitudinal structure, including the sizes and the arrangement of the layers.}
\end{figure}

The LHCf detectors are two independent sampling and imaging calorimeters, called Arm1 and Arm2. Each calorimeter is composed of two towers, made of tungsten absorbers alternated with 16 GSO scintillator layers. The lateral profile of the shower is reconstructed by imaging layers inserted at different depths. Arm1 uses four pairs of X–Y GSO bar-bundle hodoscopes \cite{suzuki2013performance}, while Arm2 uses four pairs of X-Y silicon microstrip detectors, in the first two pairs the X and Y layers are placed in contact, while in the last two pairs they are detached at different depths \cite{adriani2010construction}. The total length of the two arms is about 21 cm, equivalent to 44 radiation lengths or 1.6 interaction lengths. The calorimeter towers have transversal dimensions of 20 mm×20 mm and 40 mm×40 mm for Arm1, 25 mm×25 mm  and 32 mm×32 mm for Arm2. This configuration allows an optimal reconstruction of $\pi^{0}$ and $\eta$ mesons from the simultaneous detection of the two photons generated by their decay. The detectors are located in two regions on opposite sides of LHC Interaction Point 1 (IP1), at a distance of 141.05 m from IP1. In these regions, called Target Neutral Absorbers (TAN), the beam vacuum chamber makes a Y-shaped transition from a single beam tube facing the IP1 to two separate beam tubes joining the arcs of LHC. In this position the LHCf experiment accesses to the measurement of the high-energy neutral particle flux produced by hadronic collisions with a pseudorapidity $\left| \eta \right|>8.4$ while charged particles directed toward the LHCf detector positions are wiped out by the D1 dipole magnet. The performances of the detector were evaluated during beam tests at the CERN Super Proton Synchrotron (SPS) \cite{makino2017performance,kawade2014performance}. Concerning the reconstruction of photons, the estimated energy and position resolution values of the LHCf-Arm2 detector are better than 3\% and 40 \textmu m, respectively, for photons with energy above 200 GeV \cite{makino2017performance}. More details on the scientific goals and the performance of the LHCf experiment are discussed in previous reports \cite{makino2017performance,kawade2014performance,adriani2006lhcf}. For this analysis, we used only the LHCf-Arm2. A schematic representation of this detector is illustrated in Figure \ref{fig:lhcf_det}.

\section{\label{sec:cond}Summary of data taking conditions and Monte Carlo simulations methodology}
\subsection{\label{sec:level1}Data taking conditions}
The two experimental datasets used for this analysis were acquired by a special LHCf run on June 12th-13th, 2015, corresponding to LHC Fill 3855. During this dedicated low luminosity fill, 29 bunches collided at IP1 with a half crossing angle of 145 \textmu rad and a $\beta^{*}$ of 19 m. Other 6 and 2 non-colliding bunches circulated in the clockwise and counter-clockwise beams, respectively, which were used to estimate the background due to interactions between the produced particles and the residual gas molecules in the beam pipe. The first dataset was taken from 22:32 to 1:30 (CEST), with an instantaneous luminosity of $L=(3-5)\times10^{28}$ cm\( ^{-2} \)s\( ^{-1} \) (as measured by the ATLAS experiment \cite{collaboration2016luminosity}) and an average number of collisions per bunch crossing $\mu$ in the range of [0.007, 0.012]. The second dataset was taken from 1:40 to 12:10 (CEST), with an instantaneous luminosity of $L=(13-17)\times10^{28}$ cm\( ^{-2} \)s\( ^{-1} \)\cite{collaboration2016luminosity} and an average number of collisions per bunch crossing $\mu$ in the range of [0.03, 0.04]. Considering the LHCf data acquisition live time, the integrated luminosities were estimated to be 0.194 nb\( ^{-1} \) and 1.938 nb\( ^{-1} \) for the first and the second dataset, respectively. Both datasets were used in this analysis for a total of 8.4 million triggered events. 
\subsection{\label{sec:mc_sim}Monte Carlo simulations}
Monte Carlo (MC) simulations were used in this work to perform several analysis tasks. The full MC simulations with the same configuration of the LHCf consist of three steps: (1) hadronic interaction of proton-proton collisions at $\sqrt{s}=13$ TeV at IP1, (2) transport of produced particles from IP1 to LHCf-Arm2 location and (3) interactions with the detector. The three parts of the simulation were performed using simulation packages Cosmos 7.633 \cite{cosmos} and EPICS 9.15 \cite{epics}. The full simulation has been performed starting from the events generated by two hadronic interaction models, QGSJET II-04 \cite{ostapchenko2011monte} and EPOS-LHC \cite{pierog2015epos}. The interaction of the collision products with the detector was simulated using DPMJET 3.04 \cite{bopp2008antiparticle}. As described in Sections \ref{sec:corr_intro}-\ref{sec:err_intro}, the two full simulation datasets were used to calculate some correction factors and systematic uncertainties. The full QGSJET II-04 simulation was also used to compute the energy-dependent cut function for particle identification. To compare our results with hadronic interaction model predictions, we used the CRMC package \cite{crmc}, which acts as a frontend for the models under consideration. In this way, we generated a large simulation data set for the models that are commonly used for air shower simulations: QGSJET II-04, EPOS-LHC, SIBYLL 2.3 \cite{riehn2015new} and DPMJET 3.06. In all the four cases, we considered the particles directly produced in the collisions or by the decay of unstable particles with $c\cdot\tau<1$ cm. The number of events and the inelastic cross-section of each model is listed in Table \ref{tab:table1}. More details about the simulation's methodology can be found in \cite{adriani2012measurement}.
\begin{table}[t]
\centering
\begin{tabular}{|l|c|c|c|c|}
\hline
\textrm{Model} &
\textrm{QGSJET} &
\textrm{EPOS} &
\textrm{SIBYLL} &
\textrm{DPMJET}\\
\hline
$\sigma_{inel}$ [mb] & 80.17 & 78.98 & 79.86 & 80.14\\
$N_{ev}$ & 9.96$\cdot$$10^{7}$ & 9.90$\cdot$$10^{7}$ & $10^{8}$ & $10^{8}$\\\hline
\end{tabular}
\caption{\label{tab:table1}%
Total inelastic cross-section $\sigma_{inel}$ for p-p collisions at $\sqrt{s}=13$ TeV and the number of events ($N_{ev}$) for each hadronic interaction model used for the comparison with experimental data (version number is omitted).
}
\end{table}
\section{\label{sec:ana_frame}Analysis framework}
\subsection{\label{sec:ev_rec_and_sel}$\eta$ event reconstruction and selection}
\begin{figure}[t]
\centering
\includegraphics[width=0.8\linewidth]{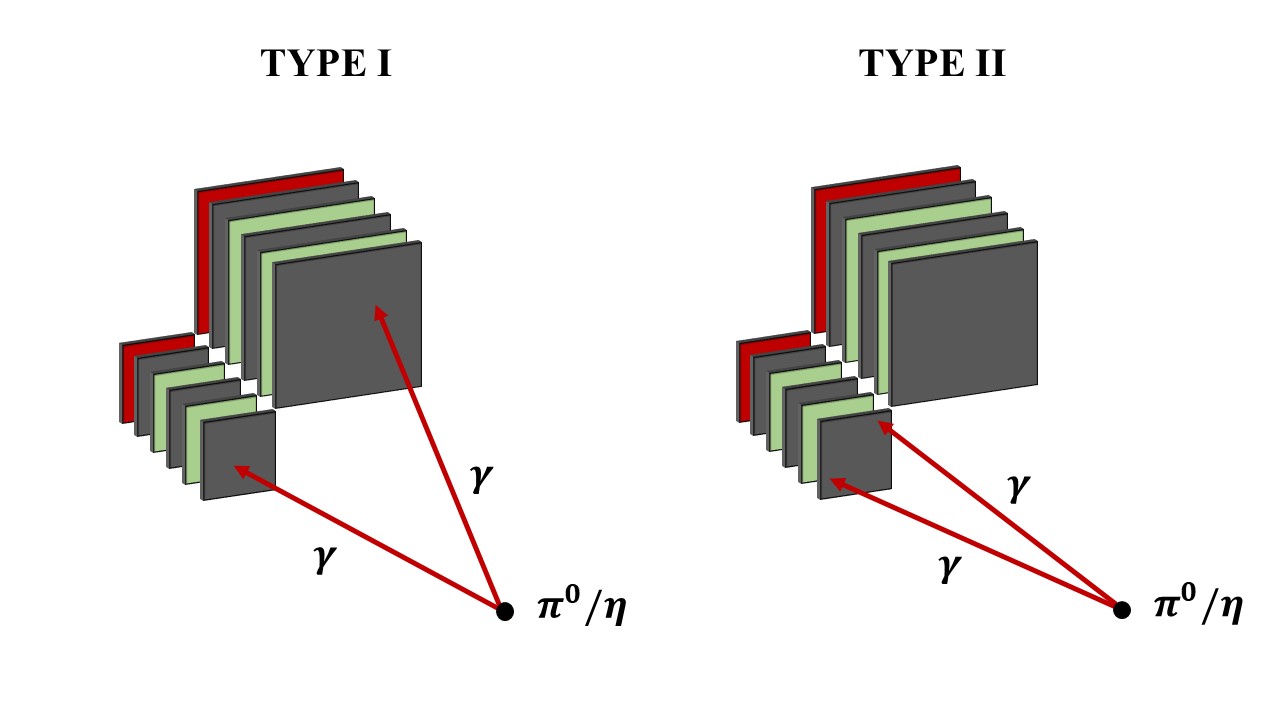}
\caption{\label{fig:type}Schematic representation of a $\pi^{0}$ or $\eta$ decay observed by the LHCf-Arm2 detector. Photons can enter in two different towers (Type I event, left panel) or both in the same tower (Type II event, right panel). The LHCf-Arm2 representation is not on scale.} 
\end{figure}

The LHCf-Arm2 detector can identify $\eta$ mesons by reconstructing the two photons produced in the decay $\eta\rightarrow\gamma\gamma$ (B.R. 39.41 \% \cite{particle2022review}). The two photons can enter one in each detector tower (Type I events, left panel of Figure \ref{fig:type}) or both in the same tower (Type II events, right panel of Figure \ref{fig:type}). The reconstruction and selection algorithms for $\eta$ mesons are similar to that developed for Type I $\pi^{0}$ analysis \cite{adriani2012measurement, adriani2016measurements}. Type II $\eta$ studies however cannot be carried out in this analysis since the acceptance for this type of event in the LHCf-Arm2 detector is very small. We do not have enough events in the datasets to obtain a significant result for Type II, hence we limit our analysis to Type I events. The $\eta$ mesons are produced in the collisions decays very close to the interaction point IP1. Indeed we calculate the opening angle $\theta$ from the transverse distance between photon impact points at the LHCf-Arm2 detector assuming that the decay happened at the IP1 (about 141.05 m from the detector). As a consequence, the opening angle is very small and constrained by $\theta\leq0.6$ mrad. We also reconstructed the kinematic variables of $\eta$ (energy, $p_{T}$ and $p_{z}$) by using the energies and positions of the photons hitting the calorimeter. The $\eta$ inclusive production rate has been expressed in terms of the Feynman-x variable, which was calculated as $x_{F}=2p_z/\sqrt{s}$. Despite interesting information about scaling laws can be extracted by looking at the $\eta$ meson $x_{F}$ distribution in several transverse momentum $p_{T}$ bins, due to the limited statistics we were able to extract such distribution for only one bin, with $0.0$ GeV/c $\leq p_t<1.1$ GeV/c. The data analysis algorithm consists of five steps: hit position reconstruction, energy reconstruction, single photon identification, $\eta$ reconstruction and background subtraction.
\subsubsection{\label{sec:hit_pos_rec}Hit position reconstruction}
The transverse position of particles hitting the LHCf-Arm2 detector is determined using the lateral profile distribution recorded by the silicon microstrip position-sensitive layers. Using an algorithm based on the TSpectrum class \cite{morhavc1997background} of the ROOT analysis framework \cite{brun1997root}, it is possible to separate events with a single particle hitting the tower (single-hit events) from that with more than one particle in the same tower (multi-hit events). Single-hit events present only one peak in the lateral profile distribution, while several peaks are identified for multi-hit events. Multi-hit events are rejected from the analysis. The loss of $\eta$ events due to the multihit cut is corrected, as explained in Section \ref{sec:multi_cor}. The lateral distributions are fit to superimposed Lorentzian functions (one for each peak) to precisely estimate the shower peak position, height, and width. Particles with hit position within 2 mm from the edges of the two towers are excluded to avoid significant effects due to the lateral shower's leakage. 
\subsubsection{\label{sec:level1}Energy reconstruction}
The energy of photons is reconstructed from the deposited energy in the calorimeter layers. Energy deposits are converted from the charge information using the conversion coefficients calculated at the SPS beam tests \cite{makino2017performance}. The deposited energy is computed using the sum of the releases from the 2nd to 13th layer and is corrected for the light yield efficiency of the scintillators and the leakage effects \cite{adriani2008lhcf}. Then, using an empirical polynomial function, we convert the total deposited energy into the primary energy. Events with energy below 200 GeV are not considered in the analysis to reject the particles produced in the beam pipe interactions and avoid uncertainties due to the trigger inefficiency.
\subsubsection{\label{sec:pid}Single photon identification}
The particle identification (PID) algorithm used for this analysis aims to separate photons from neutral hadron background, mainly due to neutrons. We perform the selection by using one particular shower feature, the longitudinal distance measured from the first calorimeter layer to the depth where the total energy deposition is 90\% of the total shower energy deposition. This variable is called $L_{90\%}$ and is expressed in units of radiation length [$X_{0}$]. In a previous study \cite{adriani2012measurement}, we demonstrated that $L_{90\%}$ has a strong discrimination power to distinguish between pure electromagnetic showers and the background showers produced by hadrons. PID criteria are expressed as a function of the energy of the particles hitting the calorimeter $f_{L_{90\%}}(E)$, in order to impose a constant selection efficiency of 90\% in the whole photon energy range. The two functions, one for each tower, are calculated using the full QGSJET II-04 model simulation. To calculate the cut functions, we applied all the single-photon selection criteria used for the analysis, described in the previous sections and listed in Table \ref{tab:table2}. Using the MC truth of the simulation, we considered only photon pairs produced in the $\eta$ decays. This is necessary to maximize the selection of $\eta$ since most of the photons that pass the imposed criteria come from the decay of $\pi^{0}$s, which possess very different kinematics than $\eta$. The residual hadron contamination is removed during the background subtraction procedure described in Section \ref{sec:bkg_sub}. The selection inefficiency is corrected in Section \ref{sec:sel_cor}, while the uncertainty related to the PID is calculated in Section \ref{sec:pid_err}.
\begin{table}[b]
\centering
\begin{tabular}{|l|l|}
\hline
Event type & Type I \\
Number of hits & single-hit for each tower \\
Incident position & Within 2 mm from the edge of calorimeter \\
Energy threshold & $E>200$ GeV\\
PID & Photonlike [$L_{90\%}$  $<$ $f_{L_{90\%}}(E)$]\\\hline
\end{tabular}
\caption{\label{tab:table2}%
List of single-photon selection criteria.
}
\end{table}
\subsubsection{\label{sec:bkg_sub}$\eta$ reconstruction and background subtraction}
\begin{figure}[t]
\includegraphics[width=0.95\linewidth]{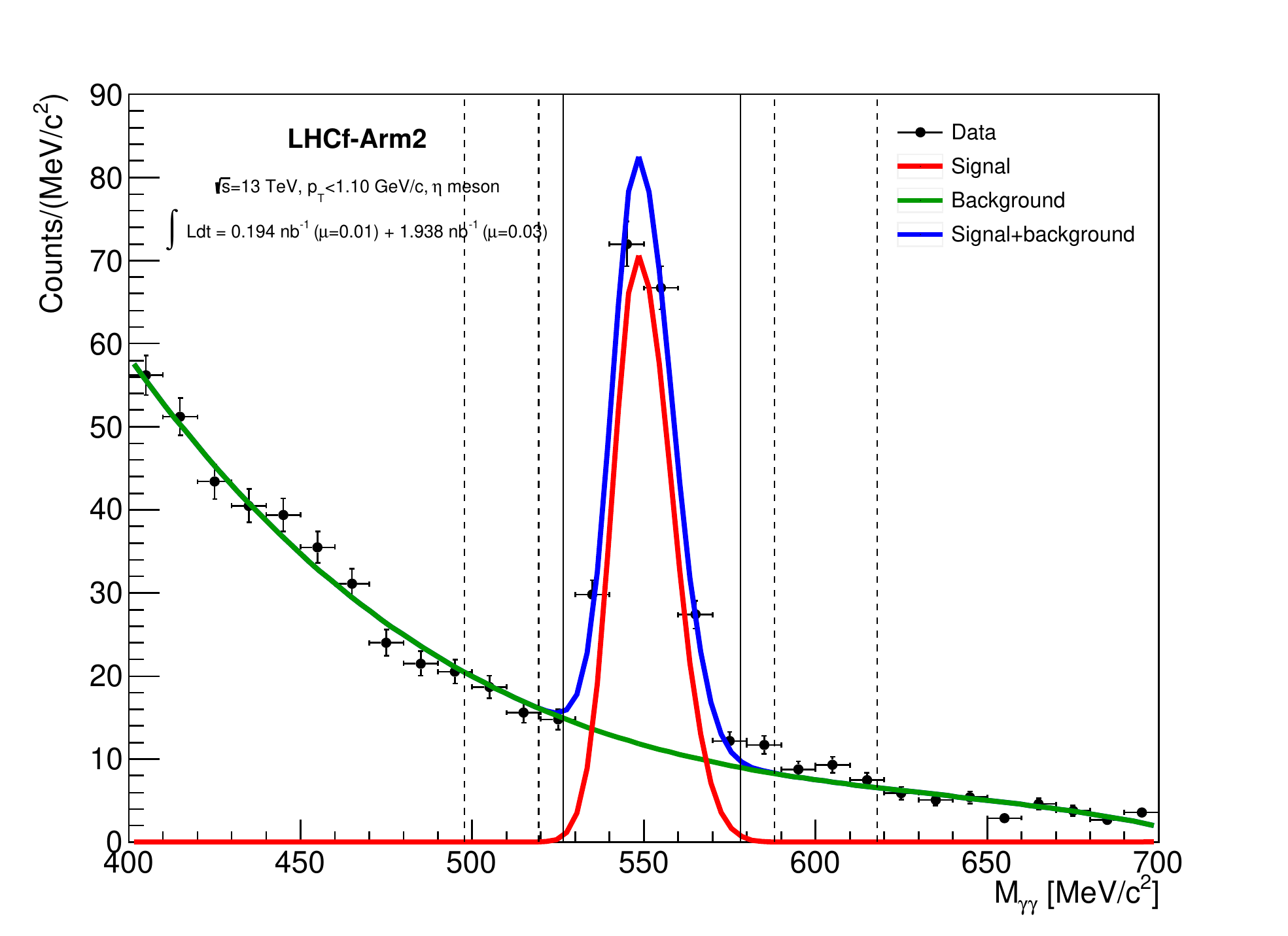}
\caption{\label{fig:Mgg} Di-photon invariant mass distribution reconstructed using the LHCf-Arm2 detector for pairs with $p_{T}<$ 1.1 GeV/c. The blue solid line shows the result of the composite fit on the distribution, obtained by adding the signal fit distribution (asymmetric Gaussian function, red line) and the background fit distribution (third-order Chebyshev polynomial function, green line). Solid and dashed vertical lines indicate the signal and the two background windows, respectively.}
\end{figure}
 Candidates $\eta$ mesons are selected by looking at the characteristic peak in the di-photon invariant mass spectrum around the $\eta$ rest mass. We use the reconstructed energy and position of selected photon pairs to compute the invariant mass $M_{\gamma\gamma}$ according to the formula:
\begin{equation}
M_{\gamma\gamma} = \sqrt{2E_{1}E_{2}(1-cos\theta)},
\end{equation}
where $E_{1}$ and $E_{2}$ are the energies of the two photons, and $\theta$ is the opening angle in the laboratory reference system. The peak of the distribution occurred at $M_{\gamma\gamma}=533.3\pm1.1$ MeV/c\( ^{2} \). The world averaged $\eta$ rest mass is $M_{\eta}=547.86$ MeV/c\( ^{2} \) \cite{particle2022review}, so a shift of $(-2.65\pm0.20)$\% is present in our data. We verified that a compatible shift was also present in the invariant mass peak associated with the decay of $\pi^{0}$ into two photons hitting the two different towers of Arm2, and we found a consistent value of $(-2.57\pm0.04)$\%. As motivated in Section \ref{sec:en_scal_err}, we applied an artificial shift of $+2.65$\% to the energies of single photons to correct the peak position according to the reference value. After this correction, the peak became concentrated around the value $M_{\gamma\gamma}=548.1\pm1.1$ MeV/c\( ^{2} \).\\ Since the $\eta$ statistic was very low in the analysed dataset, we could not extract the $x_F$ distribution and remove the combinatorial background using a template fit for each $x_F$ bin. We, therefore, decided to use a sideband method \cite{adriani2012measurement}. First, we performed a fit of the distribution using a composite model made by the sum of an asymmetric Gaussian function for the signal component and a third-order Chebyshev polynomial function for the background component. The expected mean $\bigl \langle m \bigr \rangle$ and the $1\sigma$ deviations ( $\sigma_{l}$ for the left side and $\sigma_{r}$ for the right side) have been used to define the signal region within [$\bigl \langle m \bigr \rangle-3\sigma_{l}, \bigl \langle m \bigr \rangle+3\sigma_{r}$] and two background regions, within [$\bigl \langle m \bigr \rangle-7\sigma_{l}, \bigl \langle m \bigr \rangle-4\sigma_{l}$] and [$\bigl \langle m \bigr \rangle+4\sigma_{r}, \bigl \langle m \bigr \rangle+7\sigma_{r}$]. The invariant mass distribution, the composite fit, the signal region and the two background regions are displayed in Figure \ref{fig:Mgg}. The background component in the signal region was estimated by scaling the sum of the $x_F$ distributions in the background regions for the ratio between the integrals of the Chebyshev polynomial function in the signal and background regions. Then it was subtracted to the signal region $x_{F}$ distribution. Using this method we found about 1500 $\eta$ mesons in the dataset. The uncertainty of the background subtraction method was estimated using the full reconstructed QGSJET II-04 simulation, as described in Section \ref{sec:bkg_sub_err}. 
\subsection{\label{sec:corr_intro}Corrections}
\begin{figure}[t]
\includegraphics[width=0.95\linewidth]{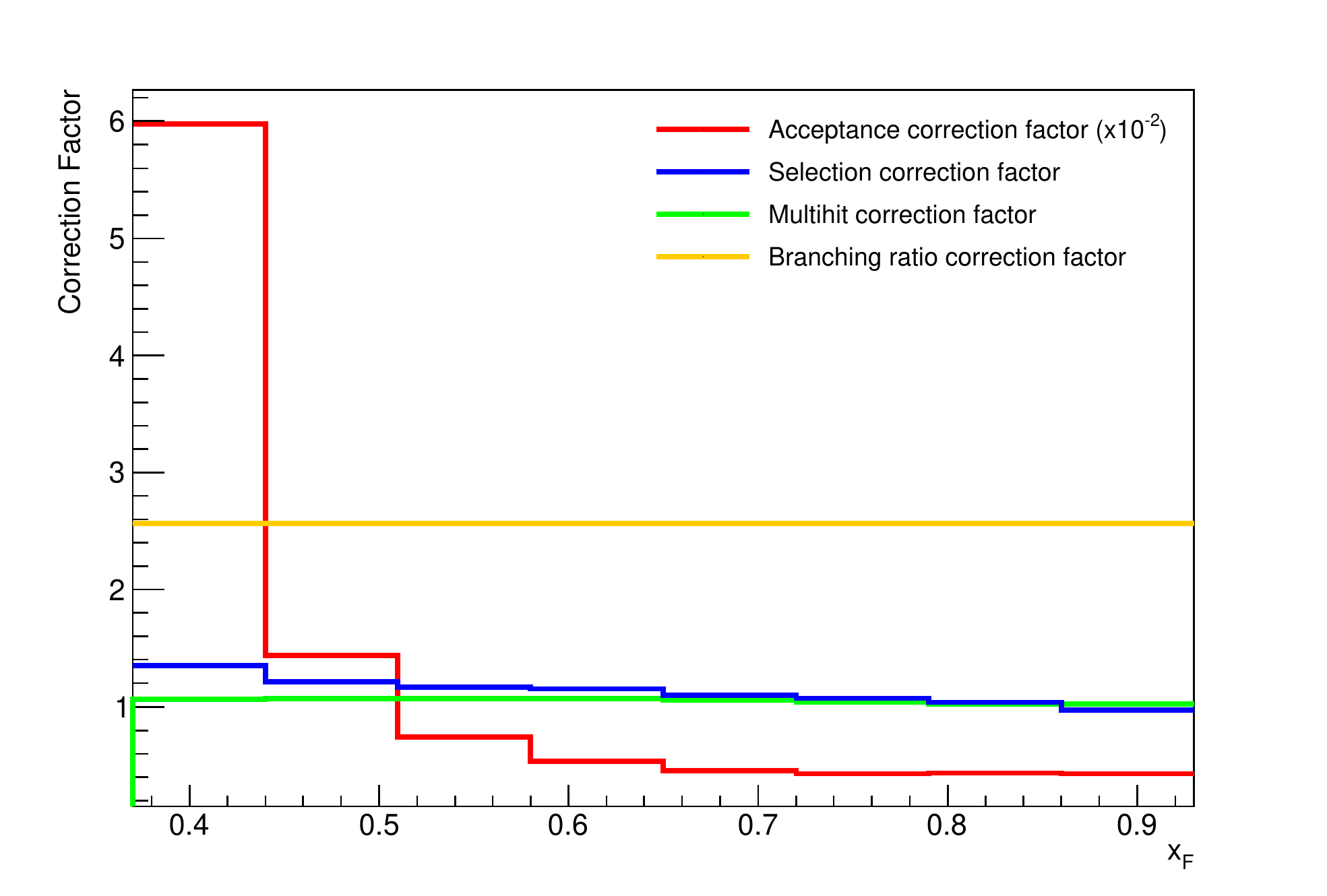}
\caption{\label{fig:corr} Corrections for experimental effects in the LHCf-Arm2 detector, applied to the final $x_{F}$ spectrum. The acceptance correction is scaled for a factor $10^{-2}$.}
\end{figure}

The corrections applied to the $x_F$ spectrum of $\eta$ are discussed in this Section. Each correction is described in Sections \ref{sec:sel_cor}-\ref{sec:multi_cor}. Figure \ref{fig:corr} shows distributions of correction factors as a function of $x_{F}$.
\subsubsection{\label{sec:sel_cor}Selection correction}
First, the signal distribution has to be corrected for the selection inefficiency and the smearing effects. Both effects are corrected simultaneously using the fully reconstructed simulation based on QGSJET II-04 and EPOS-LHC. For both models, we took the ratio between the $\eta$ candidate $x_F$ distribution, obtained using the same reconstruction algorithm as experimental data, and the true $x_F$ distribution of $\eta$ mesons. The final correction factor is estimated from the average of the values obtained from the two models. To consider the differences between the models, we add a systematic error related to this correction, described in Section \ref{sec:mc_rel_err}.

\subsubsection{\label{sec:level1}Acceptance and branching ratio correction}
\begin{figure}[]
\includegraphics[width=\linewidth, height=9cm]{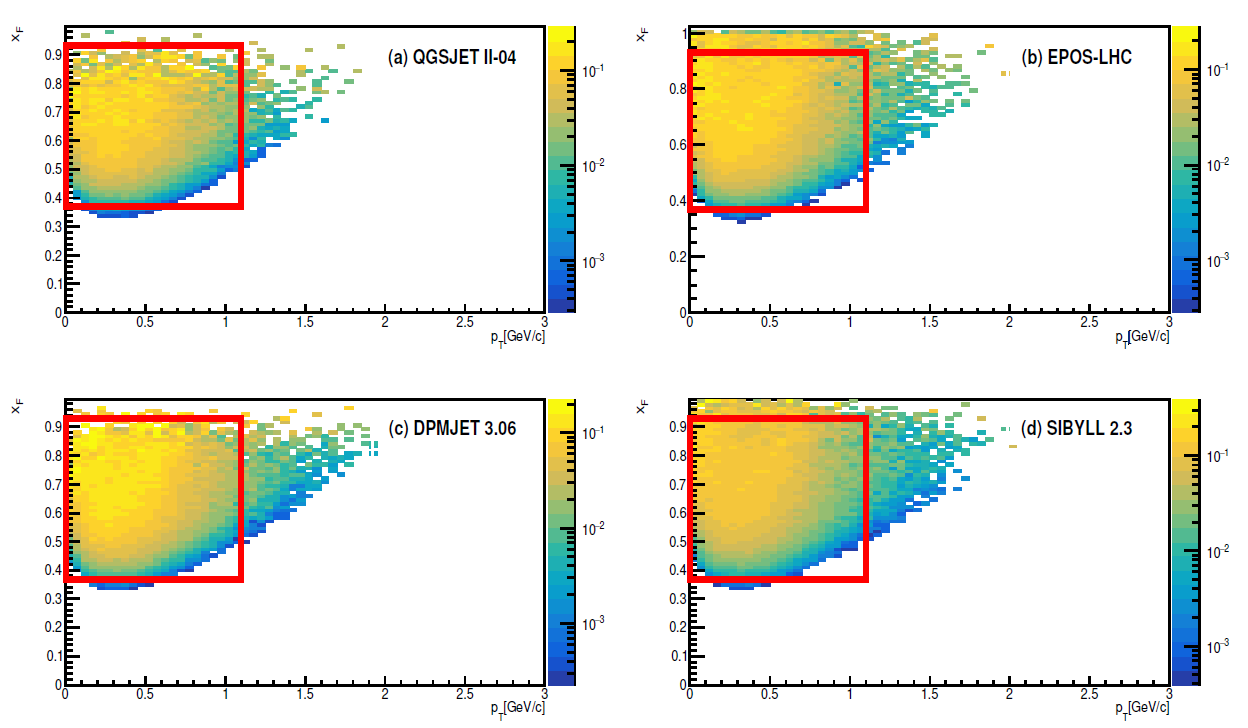}
\caption{\label{fig:acc_map}Acceptance maps of $\eta$ detection by the
LHCf-Arm2 detector in $p_{t}-x_{F}$ phase space for the hadronic interaction models considered in the analysis. Position and energy threshold cuts are taken into account. The red boxes indicate the $p_{t}-x_{F}$ analysis region of this analysis.}
\end{figure}

 Second, we corrected the signal distribution for the limited aperture of the LHCf-Arm2 detector since it does not cover the full $2\pi$ azimuthal angle. To estimate the acceptance correction factors, we used a toy MC  simulation based on the predictions from four hadronic interaction models, QGSJET II-04, EPOS-LHC, DPMJET 3.06 and SIBYLL 2.3.
 For each model, we generated an $\eta$ meson $p_{t}-x_{F}$ phase space, then we simulated the decay $\eta\rightarrow\gamma\gamma$ and computed the $p_{t}-x_{F}$ phase space for the particles hitting the LHCf-Arm2 detector, also considering the single-photon selection criteria on energy and position listed in Table \ref{tab:table2}. The geometrical acceptance efficiency was calculated as the ratio of accepted $\eta$ mesons divided by the distribution of all generated $\eta$ mesons. The four models provide different predictions of the acceptance efficiency due to the different $p_t-x_F$ spectrum shapes inside the $x_F$ bins used in the analysis, as shown in Figure \ref{fig:acc_map}, where the acceptance maps of the models considered in the analysis are displayed. The final map applied to the data was obtained by averaging the results of the models in the analysis $p_{t}-x_{F}$ region, defined by $p_{T}<1.1$ $GeV/c$ and $0.37\leq x_{F}\leq0.93$. The analysis region is indicated by red boxes in the acceptance maps shown in Figure \ref{fig:acc_map}. A systematic uncertainty was calculated using the method described in Section \ref{sec:mc_rel_err} to account for the differences between these models. 
The branching ratio of the $\eta$ decay into two photons is $39.41\%$. We also corrected this inefficiency by applying a constant factor to the signal distribution in the whole $x_{F}$ range.
\subsubsection{\label{sec:multi_cor}Multihit correction}
Last, the inefficiency due to the multi-hit rejection was corrected. This inefficiency is due to the loss of $\eta$ mesons when cutting the events with more than one particle per tower of the LHCf-Arm2 detector, described in Section \ref{sec:hit_pos_rec}. The multi-hit correction factor is defined for each bin $i$ of $x_F$ as:
\begin{equation}
f_{multihit}^{i} = \frac{N^{i}_{multihit}+N^{i}_{singlehit}}{N^{i}_{singlehit}}.
\end{equation}
where $N_{multihit}^{i}$ is the number of multihit events while $N_{singlehit}^{i}$ is the number of single-hit events, both for each bin $i$ of $x_F$. Two corrections were calculated using the results of the full simulations of QGSJET II-04 and EPOS-LHC. To account for the differences between the two models, the signal distribution was corrected using the average of the correction factors obtained from them, and an additional systematic error was calculated according to the method reported in Section \ref{sec:mc_rel_err}. 
\subsection{\label{sec:err_intro}Systematic uncertainties}
\begin{figure}[t]
\includegraphics[width=0.95\linewidth]{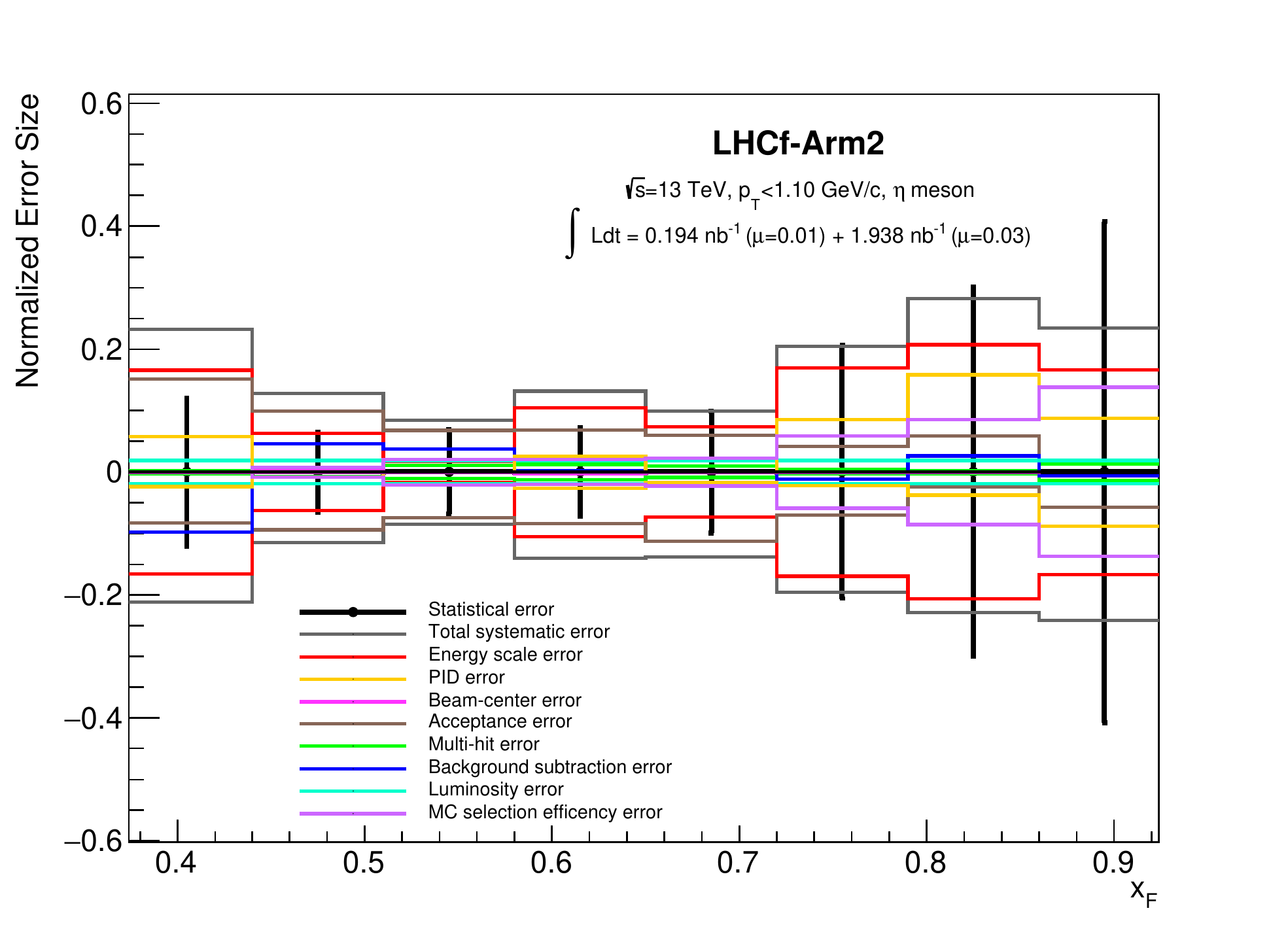}
\caption{\label{fig:err} Relative systematical uncertainties of the $\eta $ production cross section measured with the LHCf-Arm2 detector. Coloured lines refer to the single source of error  while the grey line indicates the total systematic uncertainty, obtained by summing the contribution of each source in quadrature. Black markers indicate statistical errors.}
\end{figure}

The estimation of the total uncertainty on the $x_{F}$ distribution of $\eta$ is discussed in this Section. A description of each source of error is given in sections \ref{sec:en_scal_err}-\ref{sec:mc_rel_err}. The total systematic error is calculated by quadratically summing the contribution of each source. Figure \ref{fig:err} shows the estimated systematic uncertainties as a function of $x_{F}$.

\subsubsection{\label{sec:en_scal_err}Energy scale}
As discussed in Sections \ref{sec:bkg_sub}, systematic shifts in the invariant mass peaks of $\pi^{0}$ and $\eta$ were found with respect to the world averaged values of the rest mass of the two particles. The amount of the discrepancies were $-2.57\pm0.04$\% and $-2.65\pm0.20$\% for $\pi^{0}$ and $\eta$, respectively. The two values were consistent within the errors and were also compatible with the uncertainty on the absolute energy scale, calculated using beam test data at SPS, and equal to $\pm 2.7$\% \cite{tiberio2018measurement}. The two peaks were restored to the correct position by increasing the energies of individual photons by $+2.65$\%. In order to assess the uncertainty on the absolute energy scale, we decided to check the stability of this value as a function of the energy. This was done for each tower by considering Type II events, which release all their energy in a single tower. Since Type II $\eta$ are outside detector acceptance, we used Type II $\pi^0$ for this purpose. In both towers, the stability of Type II $\pi^0$ invariant mass throughout the energy range was within 1\%, which was conservatively assumed as uncertainty on the absolute energy scale. The systematic errors on the energy scale were then obtained by producing two $x_{F}$ distributions artificially scaling the single-photon energies by $+1$\% and $-1$\%, respectively, and taking the variation from the unscaled spectrum as the estimation of the systematic uncertainty. The upper and lower error bands were conservatively symmetrized by assigning to both the maximum of the two values for each bin of $x_F$.
\subsubsection{\label{sec:pid_err}PID}
The systematic uncertainty associated with the particle identification method used in this analysis was calculated for every bin of the $x_{F}$ distribution by comparing the spectra obtained with different PID criteria. Two additional $L_{90\%}$ cut functions were calculated in the same way described in Section \ref{sec:pid} but using different values of the required efficiency, 85\% and 95\% instead of 90\%. These limits were chosen to maintain the product of efficiency and purity above 75\% in the full energy range. The whole analysis was then repeated with these different functions and the PID error was estimated using the relative deviation from the original distribution.
\subsubsection{\label{sec:level1}Beam-center stability}
The beam-center was calculated by fitting the hit position of high-energy hadrons with a two-dimensional Gaussian function, since these particles are very collimated to the beam axis. The beam-center position was used in this work to define the analysis region, so the uncertainties on its determination affect the final distribution. To account for this effect, a systematic error was estimated by shifting the position of the beam-center of $\pm 0.3$ mm in both the X and Y direction. This value is consistent with the parameter fluctuations observed run-by-run during the data taking. We compared the four obtained spectra with the original one and assigned the systematic uncertainties due to the beam-center stability as the relative deviations between them.
\subsubsection{\label{sec:level1}Luminosity}
The uncertainty on the integrated luminosity measured by ATLAS was estimated to be $\pm$1.9\%. Note that this is the only energy-independent systematic uncertainty. 
\subsubsection{\label{sec:bkg_sub_err}Background subtraction}
The uncertainty on the background subtraction method, explained in Section \ref{sec:bkg_sub}, is evaluated using the full MC simulation based on QGSJET II-04. The whole analysis procedure is applied to this MC dataset, up to the step where the spectrum in $x_{F}$ is extracted using the sideband method. Another $x_F$ spectrum is generated rejecting background events using the MC truth information from the simulation. The relative variation between these two spectra is used to estimate the error associated to the background subtraction method.
\subsubsection{\label{sec:mc_rel_err}MC related correction}
As described in Section \ref{sec:sel_cor}-\ref{sec:multi_cor}, we used MC simulation to calculate several corrections, namely selection, acceptance and multi-hit corrections. A systematic error was calculated for each correction to avoid the model dependence on the final distribution. The selection and multi-hit corrections were calculated using the full reconstructed simulations based on QGSJET II-04 and EPOS-LHC. In this case, the errors were calculated by the relative deviation between the correction values predicted by the two models. For the acceptance correction, we used the prediction from four different models used in this analysis, QGSJET II-04, EPOS-LHC, DPMJET 3.06 and SIBYLL 2.3. The corresponding errors were calculated in the most conservative way by looking at the relative maximum and minimum differences between the model predictions of the correction value and the mean of the values.
\section{\label{sec:res}Results}
\begin{figure}[t]
\includegraphics[width=0.95\linewidth]{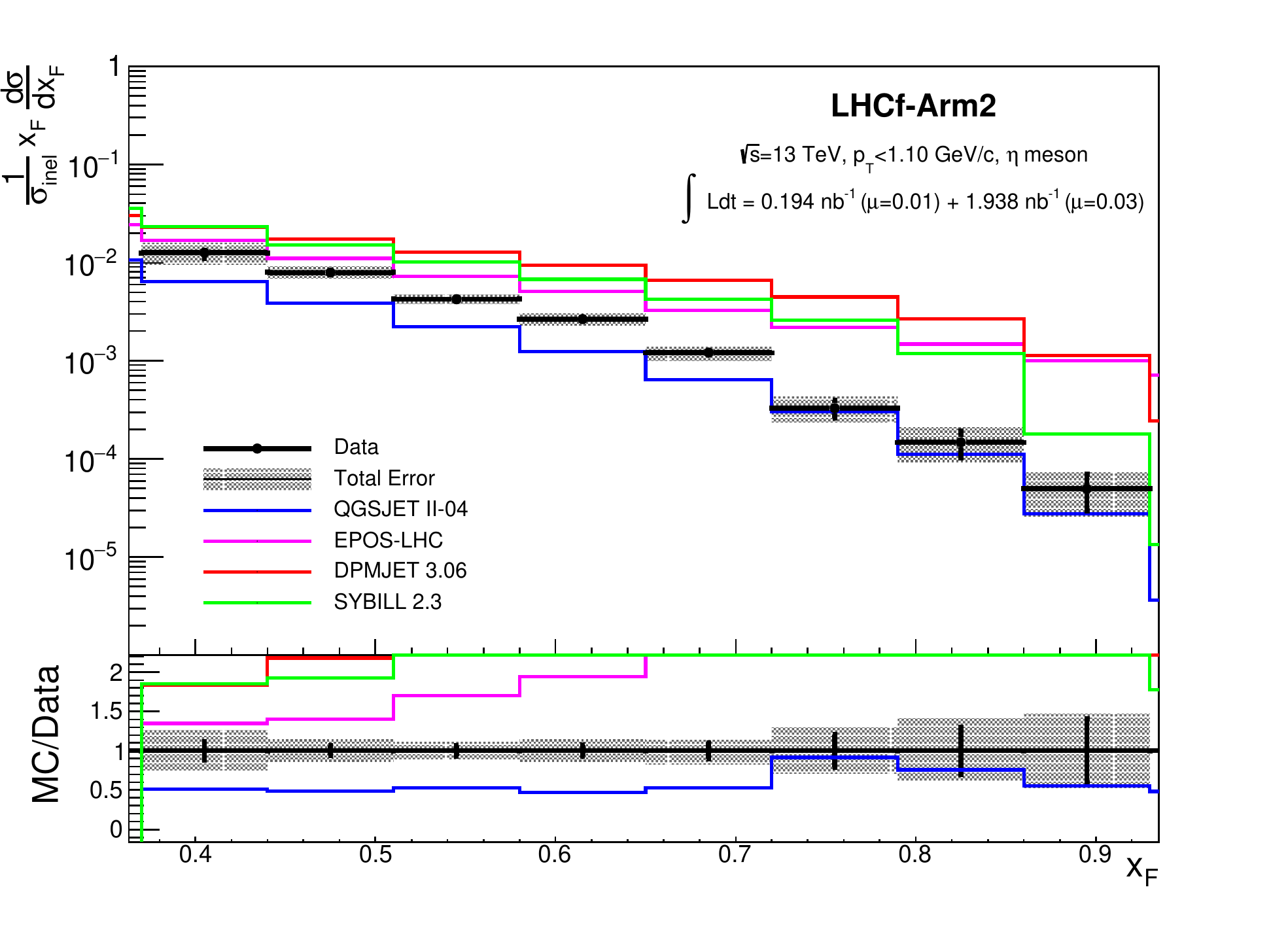}
\caption{\label{fig:fin} Inclusive $\eta$ production rate as function of $x_{F}$ in $p_{T}<$1.1 GeV/c for p-p collisions at $\sqrt{s}$ = 13 TeV, measured using the LHCf-Arm2 detector. Black markers refer to experimental data with statistical errors and grey bands refer to the total uncertainties, obtained by summing in quadrature the statistical and systematic errors. The data points are compared with the prediction of the hadronic interaction models considered in this analysis,  QGSJET II-04 (blue line), EPOS-LHC (magenta line), SIBYLL 2.3 (green line) and DPMJET 3.06 (red line).}
\end{figure}

The $x_{F}$ spectrum of $\eta$ mesons measured by the LHCf-Arm2 detector in $p_{T}<$1.1 GeV/c is presented in Figure \ref{fig:fin}. The inclusive production rate is given by the expression:
\begin{equation}
\frac{1}{\sigma_{inel}} x_{F} \frac{d\sigma}{dx_{F}}.
\end{equation}
$\sigma_{inel}$ is the inelastic cross section for proton-proton collisions at $\sqrt{s}$=13 TeV,  measured by the TOTEM experiment \cite{anelli2008totem} as $\sigma_{inel}$=79.5$\pm$1.8 mb \cite{antchev2019first}. The quantity $x_{F}$$d\sigma/dx_{F}$ is the differential cross section of  $\eta$ production, with $d\sigma=dN_{\eta}/\int Ldt$, where $dN_{\eta}$ is the number of $\eta$ mesons and $\int Ldt$ the integrated luminosity of the dataset. The black error bars in Figure \ref{fig:fin} represent the statistical uncertainties, while the grey shaded areas are the total uncertainties obtained by summing statistical and systematic errors in quadrature. The inclusive $\eta$ production rate values for each bin of $x_{F}$ and the total uncertainties are shown in Table \ref{tab:table_app_a}. In Figure \ref{fig:fin} the $x_F$ spectra predicted by several hadronic interaction models, QGSJET II-04, EPOS-LHC, DPMJET 3.06 and SIBYLL 2.3, and their ratio to experimental data are also reported. QGSJET II-04 shows the best agreement with LHCf data among the models tested in this analysis, especially for $x_{F}>0.7$, but about a factor 2 of difference is found at lower $x_{F}$. The other three models, EPOS-LHC, DMPJET 3.06 and SIBYLL 2.3 predict higher production rates and a harder spectrum with respect to experimental data in the whole $x_{F}$ range. The values of the ratios for each model and for each $x_{F}$ bin are reported in Table \ref{tab:table_app_b}.
\section{\label{sec:concl}Conclusions}
The LHCf experiment measured the inclusive production rate of $\eta$ mesons in proton-proton collisions at $\sqrt{s}=13$ TeV in $p_{T}<1.1$ GeV/c. About 1500 candidate $\eta$ mesons were found in the data set considered for this analysis; this is both the first measurement of $\eta$ mesons from LHCf and the first time a particle containing strange quarks has been observed in the very forward region for high-energy collisions. The result was compared with the prediction of several hadronic interaction models, QGSJET II-04, EPOS-LHC, DMPJET 3.06 and SIBYLL 2.3. None of the models reproduces the experimental distribution in the whole $x_{F}$ range. QGSJET II-04 shows the best agreement, but significant differences are present at low $x_{F}$. The other models predict an overall higher production rate than the experimental data. The large experimental uncertainties in this analysis, due to the low accumulated statistics, will be improved by the new LHCf data collected during LHC RUN III, in which an increase in eta meson statistics by about a factor of ten is expected \cite{adriani2019lhcf}.

\begin{acknowledgments}
We thank the CERN staff and the ATLAS Collaboration for their essential contributions to the successful operation of LHCf. We are grateful to S. Ostapchenko for useful comments about QGSJET II-04 generator and to the developers of CRMC interface tool for its implementation. This work was supported by several institutions in Japan and in Italy: in Japan, by the Japanese Society for the Promotion of Science (JSPS) KAKENHI (Grant Numbers JP26247037, JP23340076) and the joint research program of the Institute for Cosmic Ray Research (ICRR), University of Tokyo; in Italy, by Istituto Nazionale di Fisica Nucleare (INFN) and by the University of Catania (Grant Numbers UNICT 2020-22 Linea 2). This work took advantage of computer resources supplied by ICRR (University of Tokyo), CERN and CNAF (INFN).
\end{acknowledgments}

\appendix

\section{Cross section tables}
Inclusive $\eta$ production rate values as function of $x_{F}$ in $p_{T}<$1.1 GeV/c measured by LHCf-Arm2 in p-p collisions at $\sqrt{s}$ = 13 TeV are summarized in Table \ref{tab:table_app_a}. The ratios of the $\eta$ production rate of hadronic interaction models to data are summarized in Table \ref{tab:table_app_b}.

\begin{table}[h]
\centering

\begin{tabular}{|c|c|}
\hline
\textrm{$x_{F}$ range} &
\textrm{$(x_{F}/\sigma_{inel})(d\sigma/dx_{F})$} \\\hline
[0.37-0.44] & ($1.26^{+0.33}_{-0.31}$) $\times$ $10^{-2}$\\\relax
[0.44-0.51] & ($0.79^{+0.11}_{-0.11}$) $\times$ $10^{-2}$\\\relax
[0.51-0.58] & ($4.25^{+0.46}_{-0.46}$) $\times$ $10^{-3}$\\\relax
[0.58-0.65] & ($2.64^{+0.40}_{-0.41}$) $\times$ $10^{-3}$\\\relax
[0.65-0.72] & ($1.21^{+0.17}_{-0.21}$) $\times$ $10^{-3}$\\\relax
[0.72-0.79] & ($3.30^{+0.96}_{-0.94}$) $\times$ $10^{-4}$\\\relax
[0.79-0.86] & ($1.15^{+0.60}_{-0.56}$) $\times$ $10^{-4}$\\\relax
[0.86-0.93] & ($0.49^{+0.23}_{-0.24}$) $\times$ $10^{-4}$\\\hline
\end{tabular}
\caption{\label{tab:table_app_a}Inclusive $\eta$ production rate for each bin of $x_{F}$ in $p_{T}<$1.1 GeV/c, measured using the LHCf-Arm2 in p-p collisions at $\sqrt{s}$ = 13 TeV. Total uncertainties are also reported.
}
\end{table}

\begin{table}[h]
\centering
\begin{tabular}{|c|c|c|c|c|}
\hline
\textrm{$x_{F}$ range} &
\textrm{QGSJET} &
\textrm{EPOS} &
\textrm{SIBYLL} &
\textrm{DPMJET} \\
&
\textrm{II-04} &
\textrm{LHC} &
\textrm{2.3} &
\textrm{3.06} \\
\hline
[0.37-0.44] & 0.51 & 1.35 & 1.85 & 1.83 \\\relax
[0.44-0.51] & 0.49 & 1.40 & 1.92 & 2.18 \\\relax
[0.51-0.58] & 0.52 & 1.71 & 2.40 & 3.03 \\\relax
[0.58-0.65] & 0.47 & 1.94 & 2.55 & 3.55 \\\relax
[0.65-0.72] & 0.53 & 2.69 & 3.48 & 5.45 \\\relax
[0.72-0.79] & 0.91 & 6.56 & 7.88 & 13.49 \\\relax
[0.79-0.86] & 0.75 & 10.06 & 8.03 & 17.90 \\\relax
[0.86-0.93] & 0.55 & 20.23 & 3.64 & 22.72 \\
\hline
\end{tabular}
\caption{\label{tab:table_app_b}Ratio of inclusive $\eta$ production rates of hadronic interaction models to data in p-p collision at $\sqrt{s}=$ 13 TeV for each $x_{F}$ bin in $p_{T}<$1.1 GeV/c.}
\end{table}

\newpage
\bibliographystyle{JHEP}
\bibliography{jhepexample}

\providecommand{\noopsort}[1]{}\providecommand{\singleletter}[1]{#1}%
\providecommand{\href}[2]{#2}\begingroup\raggedright\begin{thebibliography}{10}

\bibitem{abraham2008observation}
J.~Abraham et~al., {\it Observation of the suppression of the flux of cosmic
  rays above $4\times$ $10^{19}$ e{V}},  {\em Physical Review Letters} {\bf
  101} (2008), no.~6 061101.

\bibitem{abreu2010update}
P.~Abreu et~al., {\it Update on the correlation of the highest energy cosmic
  rays with nearby extragalactic matter},  {\em Astroparticle Physics} {\bf 34}
  (2010), no.~5 314--326.

\bibitem{abraham2010measurement}
J.~Abraham et~al., {\it Measurement of the depth of maximum of extensive air
  showers above $10^{18}$ e{V}},  {\em Physical review letters} {\bf 104}
  (2010), no.~9 091101.

\bibitem{abbasi2009measurement}
R.~Abbasi et~al., {\it Measurement of the flux of ultra high energy cosmic rays
  by the stereo technique},  {\em Astroparticle Physics} {\bf 32} (2009), no.~1
  53--60.

\bibitem{abbasi2010analysis}
R.~Abbasi et~al., {\it Analysis of large-scale anisotropy of ultra-high energy
  cosmic rays in {H}i{R}es data},  {\em The Astrophysical Journal Letters} {\bf
  713} (2010), no.~1 L64.

\bibitem{abbasi2010indications}
R.~Abbasi et~al., {\it Indications of proton-dominated cosmic-ray composition
  above 1.6 {E}e{V}},  {\em Physical Review Letters} {\bf 104} (2010), no.~16
  161101.

\bibitem{sagawa2011recent}
{Telescope Array Collaboration}, {\it Recent results from the {T}elescope
  {A}rray experiment},  in {\em AIP Conference Proceedings}, vol.~1367,
  pp.~17--22, American Institute of Physics, 2011.

\bibitem{pierre2015pierre}
{Pierre Auger Collaboration}, {\it The {P}ierre {A}uger cosmic ray
  observatory},  {\em Nuclear Instruments and Methods in Physics Research
  Section A: Accelerators, Spectrometers, Detectors and Associated Equipment}
  {\bf 798} (2015) 172--213.

\bibitem{fukushima2003telescope}
M.~Fukushima, {\it Telescope array project for extremely high energy cosmic
  rays},  {\em Progress of Theoretical Physics Supplement} {\bf 151} (2003)
  206--210.

\bibitem{aab2014depth}
A.~Aab et~al., {\it Depth of maximum of air-shower profiles at the {P}ierre
  {A}uger {O}bservatory. {I}. measurements at energies above $10^{17.8}$ e{V}},
   {\em Physical Review D} {\bf 90} (2014), no.~12 122005.

\bibitem{aab2014muons}
A.~Aab et~al., {\it Muons in air showers at the {P}ierre {A}uger {O}bservatory:
  {M}easurement of atmospheric production depth},  {\em Physical Review D} {\bf
  90} (2014), no.~1 012012.

\bibitem{abbasi2015study}
R.~Abbasi et~al., {\it Study of {U}ltra-{H}igh {E}nergy {C}osmic {R}ay
  composition using {T}elescope {A}rray’s {M}iddle {D}rum detector and
  surface array in hybrid mode},  {\em Astroparticle Physics} {\bf 64} (2015)
  49--62.

\bibitem{gribov1968reggeon}
V.~Gribov, {\it A reggeon diagram technique},  {\em Sov. Phys. JETP} {\bf 26}
  (1968), no.~2 414--423.

\bibitem{regge1959introduction}
T.~Regge, {\it Introduction to complex orbital momenta},  {\em Il Nuovo Cimento
  (1955-1965)} {\bf 14} (1959), no.~5 951--976.

\bibitem{evans2008lhc}
L.~Evans and P.~Bryant, {\it L{H}{C} machine},  {\em Journal of
  instrumentation} {\bf 3} (2008), no.~08 S08001.

\bibitem{albrecht2022muon}
J.~Albrecht et~al., {\it The {M}uon {P}uzzle in cosmic-ray induced air showers
  and its connection to the {L}arge {H}adron {C}ollider},  {\em Astrophysics
  and Space Science} {\bf 367} (2022), no.~3 1--50.

\bibitem{adriani2008lhcf}
O.~Adriani et~al., {\it The {L}{H}{C}f detector at the {C}{E}{R}{N} {L}arge
  {H}adron {C}ollider},  {\em Journal of Instrumentation} {\bf 3} (2008),
  no.~08 S08006.

\bibitem{adriani2019lhcf}
O.~Adriani et~al., {\it {LHC}f-technical proposal for the {LHC} {R}un3},  tech.
  rep., 2019.

\bibitem{suzuki2013performance}
T.~Suzuki, K.~Kasahara, K.~Kawade, T.~Murakami, K.~Masuda, T.~Sako, and
  S.~Torii, {\it Performance of very thin {$Gd_{2}SiO_{5}$} scintillator bars
  for the {LHC}f experiment},  {\em Journal of Instrumentation} {\bf 8} (2013),
  no.~01 T01007.

\bibitem{adriani2010construction}
O.~Adriani et~al., {\it The construction and testing of the silicon position
  sensitive modules for the {LHC}f experiment at {CERN}},  {\em Journal of
  Instrumentation} {\bf 5} (2010), no.~01 P01012.

\bibitem{makino2017performance}
Y.~Makino et~al., {\it Performance study for the photon measurements of the
  upgraded {LHC}f calorimeters with {$Gd_{2}SiO_{5}$} ({GSO}) scintillators},
  {\em Journal of Instrumentation} {\bf 12} (2017), no.~03 P03023.

\bibitem{kawade2014performance}
K.~Kawade et~al., {\it The performance of the {LHC}f detector for hadronic
  showers},  {\em Journal of Instrumentation} {\bf 9} (2014), no.~03 P03016.

\bibitem{adriani2006lhcf}
O.~Adriani et~al., {\it {LHC}f experiment: technical design report},  tech.
  rep., CERN, 2006.

\bibitem{collaboration2016luminosity}
{ATLAS Collaboration}, {\it Luminosity determination in $pp$ collisions at
  $\sqrt{s}=13$ {T}e{V} using the {ATLAS} detector at the {LHC}},
  \href{http://arxiv.org/abs/2212.09379}{{\tt arXiv:2212.09379}}.

\bibitem{cosmos}
K.~Kasahara, ``Cosmos home page.''
  \url{http://cosmos.n.kanagawa-u.ac.jp/cosmosHome/index.html}, 2010.

\bibitem{epics}
K.~Kasahara, ``{EPICS} home page.''
  \url{http://cosmos.n.kanagawa-u.ac.jp/EPICSHome/index.html}, 2010.

\bibitem{ostapchenko2011monte}
S.~Ostapchenko, {\it Monte {C}arlo treatment of hadronic interactions in
  enhanced pomeron scheme: {QGSJET-II} model},  {\em Physical Review D} {\bf
  83} (2011), no.~1 014018.

\bibitem{pierog2015epos}
T.~Pierog, I.~Karpenko, J.~M. Katzy, E.~Yatsenko, and K.~Werner, {\it {EPOS
  LHC}: {T}est of collective hadronization with data measured at the {CERN}
  {L}arge {H}adron {C}ollider},  {\em Physical Review C} {\bf 92} (2015), no.~3
  034906.

\bibitem{bopp2008antiparticle}
F.~W. Bopp, J.~Ranft, R.~Engel, and S.~Roesler, {\it Antiparticle to particle
  production ratios in hadron-hadron and d-{A}u collisions in the {DPMJET-III}
  monte carlo model},  {\em Physical Review C} {\bf 77} (2008), no.~1 014904.

\bibitem{crmc}
T.~Pierog, C.~Baus, and R.~Ulrich.
  \url{https://web.ikp.kit.edu/rulrich/crmc.html}.

\bibitem{riehn2015new}
F.~Riehn, R.~Engel, A.~Fedynitch, T.~K. Gaisser, and T.~Stanev, {\it A new
  version of the event generator {S}ibyll},  {\em arXiv preprint
  arXiv:1510.00568} (2015).

\bibitem{adriani2012measurement}
O.~Adriani et~al., {\it Measurement of forward neutral pion transverse momentum
  spectra for $\sqrt{s}$= 7 {T}e{V} proton-proton collisions at the {LHC}},
  {\em Physical Review D} {\bf 86} (2012), no.~9 092001.

\bibitem{particle2022review}
R.~Workman et~al., {\it Review of particle physics},  {\em Progress of
  theoretical and experimental physics} {\bf 2022} (2022), no.~8 083C01.

\bibitem{adriani2016measurements}
O.~Adriani et~al., {\it Measurements of longitudinal and transverse momentum
  distributions for neutral pions in the forward-rapidity region with the
  {LHC}f detector},  {\em Physical Review D} {\bf 94} (2016), no.~3 032007.

\bibitem{morhavc1997background}
M.~Morh{\'a}{\v{c}}, J.~Kliman, V.~Matou{\v{s}}ek, M.~Veselsk{\`y}, and
  I.~Turzo, {\it Background elimination methods for multidimensional
  coincidence $\gamma$-ray spectra},  {\em Nuclear Instruments and Methods in
  Physics Research Section A: Accelerators, Spectrometers, Detectors and
  Associated Equipment} {\bf 401} (1997), no.~1 113--132.

\bibitem{brun1997root}
R.~Brun and F.~Rademakers, {\it {ROOT}-an object oriented data analysis
  framework},  {\em Nuclear instruments and methods in physics research section
  A: accelerators, spectrometers, detectors and associated equipment} {\bf 389}
  (1997), no.~1-2 81--86.

\bibitem{tiberio2018measurement}
O.~Adriani et~al., {\it Measurement of forward photon production cross-section
  in proton-proton collisions at $\sqrt{s}$=13 {T}e{V} with the {LHC}f
  detector},  {\em Physics Letters B} {\bf 780} (2018) 233--239.

\bibitem{anelli2008totem}
G.~Anelli et~al., {\it The {TOTEM} experiment at the {CERN} large hadron
  collider},  {\em Journal of Instrumentation} {\bf 3} (2008), no.~08 S08007.

\bibitem{antchev2019first}
G.~Antchev et~al., {\it First measurement of elastic, inelastic and total
  cross-section at $\sqrt{s}$=13 {T}e{V} by {TOTEM} and overview of
  cross-section data at {LHC} energies},  {\em The European Physical Journal C}
  {\bf 79} (2019), no.~2 1--10.

\end{thebibliography}\endgroup






\end{document}